\documentclass[aps,showpacs,twocolumn,amsmath,amssymb,superscriptaddress,prl]{revtex4-1}

\usepackage[dvipdfmx]{graphicx}
\usepackage{color}

\usepackage{bm}
\usepackage{comment}

\begin{document}

\title{Role of interfacial oxidation in generation of spin-orbit torques}

\author{Satoshi Haku}
\affiliation{Department of Applied Physics and Physico-Informatics, Keio University, Yokohama 223-8522, Japan}

\author{Musha Akira}
\affiliation{Department of Applied Physics and Physico-Informatics, Keio University, Yokohama 223-8522, Japan}

\author{Tenghua Gao}
\affiliation{Department of Applied Physics and Physico-Informatics, Keio University, Yokohama 223-8522, Japan}
\affiliation{Keio Institute of Pure and Applied Science (KiPAS), Keio University, Yokohama 223-8522, Japan}

\author{Kazuya Ando}
\email{ando@appi.keio.ac.jp}
\affiliation{Department of Applied Physics and Physico-Informatics, Keio University, Yokohama 223-8522, Japan}
\affiliation{Keio Institute of Pure and Applied Science (KiPAS), Keio University, Yokohama 223-8522, Japan}
\affiliation{Center for Spintronics Research Network (CSRN), Keio University, Yokohama 223-8522, Japan}

\date{\today}

\begin{abstract}
We report that current-induced spin-orbit torques (SOTs) in heavy-metal/ferromagnetic-metal bilayers are strongly altered by the oxidation of the ferromagnetic layer near the interface. We measured damping-like (DL) and field-like (FL) SOTs for Pt/Co and Pt/Ni$_{81}$Fe$_{19}$ (Pt/Py) films using spin-torque ferromagnetic resonance. In the Pt/Co film, we found that the oxidation of the Co layer near the interface enhances both DL and FL SOTs in spite of the insulating nature of the CoO$_x$ layer. The enhancement of the SOTs disappears by inserting a thin Ti layer at the Pt/CoO$_x$ interface, indicating that the dominant source of the SOTs in the Pt/CoO$_x$/Co film is the spin-orbit coupling at the Pt/CoO$_x$ interface. 
In contrast to the Pt/CoO$_x$/Co film, the SOTs in the Pt/PyO$_x$/Py film are dominated by the bulk spin-orbit coupling. Our result shows that the interfacial oxidation of the Pt/Py film suppresses the DL-SOT and reverses the sign of the FL-SOT. The change of the SOTs can be attributed to the change of the real and imaginary parts of the spin mixing conductance induced by the insertion of the insulating PyO$_x$ layer. These results show that the interfacial oxidation provides an effective way to manipulate the strength and sign of the SOTs. 
\end{abstract}

\maketitle

\section{I. introduction}

Current-induced spin-orbit torques (SOTs) in heavy-metal/ferromagnetic-metal (HM/FM) heterostructures provide a route to realize efficient spin-based nonvolatile memories and logic devices~\cite{AndoPRL, miron2011perpendicular,Liu555, PhysRevLett.109.096602, garello2013symmetry, yu2014switching,RevModPhys.91.035004}. The SOTs are generated by bulk and interfacial spin-orbit coupling in the HM/FM structure~\cite{AndoPRL,Liu_science,miron2011perpendicular,nakayama2016rashba,fan2013observation,garello2013symmetry,PhysRevLett.109.096602,qiu2015spin,Gambardella3175,miron2010current,PhysRevB.94.214417}. The bulk spin-orbit coupling in the HM layer generates the SOTs by the bulk spin Hall effect~\cite{AndoPRL, Liu555}. The interfacial spin-orbit coupling also contributes to the SOTs through various mechanisms, such as the Rashba-Edelstein effect and interfacial spin-orbit scattering~\cite{PhysRevB.78.212405,miron2010current,miron2011perpendicular,PhysRevB.94.104420}. Of particular importance to understand the physics behind the SOT generation, as well as to realize the SOT devices, is the manipulation of the SOTs in the HM/FM structures. Since several bulk and interfacial mechanisms contribute to the SOT generation in the HM/FM structures, the magnitude and sign of the SOTs can be varied by tuning the layer thicknesses~\cite{kim2013layer} and the interface engineering of the HM/FM structure~\cite{zhang2015role}. The generation efficiency of the SOTs has been shown to be enhanced by doing HM impurities with strong spin-orbit coupling~\cite{zhu2018highly,zhu2019strong,musha2019extrinsic,chen2017tunable}.

Recent studies have revealed an alternative, effective way to control the SOTs: oxidation-level engineering of the SOTs. Although the spin-orbit coupling of oxygen is quite weak, contrary to HM impurities, the oxygen incorporation dramatically alters the SOTs in the HM/FM structures. In the HM/FM structure, the control of the oxidation level of the FM layer enables to engineer the SOTs~\cite{qiu2015spin}. The oxidation of the HM layer also provides an effective way to enhance the SOT efficiencies~\cite{demasius2016enhanced,Aneaar2250,an2018giant,AnCu,kageyama2019spin,PhysRevLett.121.017202}. The oxidation-level engineering opens a route to manipulate the SOTs through voltage-driven oxygen migration~\cite{emori2014large,Aneaar2250,mishra2019electric}. Although these studies suggest an important role of the SOTs generated at the oxidized HM/FM interfaces, the role of the oxidation of the FM layer near the interface is still unclear.

In this paper, we report the role of the interfacial oxidation in the generation of the SOTs in the HM/FM films. We measured damping-like (DL) and field-like (FL) SOTs for Pt/Co and Pt/Ni$_{81}$Fe$_{19}$ (Pt/Py) films. Our results show that the interfacial oxidation can enhance or suppress the SOTs, depending on the ferromagnetic layer. The enhancement of the SOTs appears in the Pt/Co bilayer, where the interfacial oxidation enhances the interfacial spin-orbit coupling. The situation is different in the Pt/Py bilayer, where the interfacial spin-orbit coupling is suppressed by the interfacial oxidation. In the Pt/Py film, the role of the interfacial oxidation is solely to change the real and imaginary parts of the spin mixing conductance, resulting in the suppression of the DL-SOT and enhancement of the FL-SOT.

\section{II. experimental method}

\begin{figure}[tb]
\includegraphics[scale=1.2]{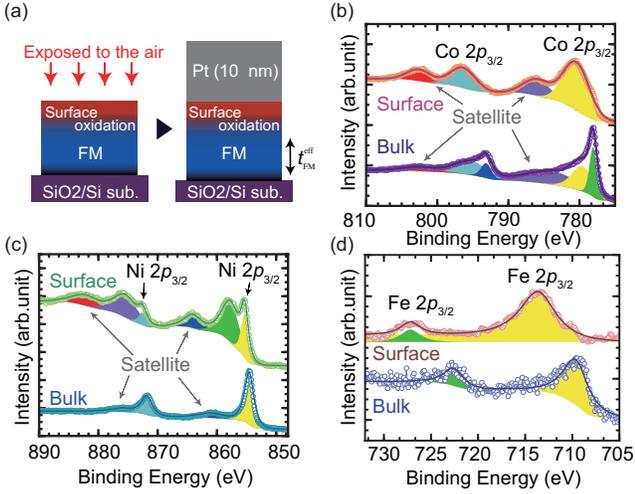}
\caption{(a) Schematic of the oxidized Pt/FM films, where the surface of the FM layer was exposed to the air for 30 minutes before the Pt-layer deposition. (b) The XPS spectra of Co 2$p$ from the surface and bulk of the surface oxidized Co film. The XPS spectra of (c) Ni 2$p$ and (d) Fe 2$p$ from the surface and bulk of the surface oxidized Py film. The symbols represent experimental data and the curves are the fitting results.
}
\label{fig2} 
\end{figure}

The effect of the interfacial oxidation on the generation of the SOTs was studied in two standard systems, Pt/Co and Pt/Py films. The Co and Py layers with the thickness of $t_{\rm FM}$ were respectively deposited on thermally oxidized Si substrates by sputtering, and subsequently exposed to the air for 30 minutes, which results in the formation of surface oxidized layers: CoO$_{x}$ and PyO$_{x}$. On the surface of the oxidized FM layer, a 10-nm-thick Pt layer was sputtered (see Fig~\ref{fig2}(a)). For comparison, non-oxidized Pt/Co and Pt/Py films were also fabricated by sputtering without breaking the vacuum during the deposition.

The formation of the CoO$_{x}$ and PyO$_{x}$ layers on the surface of the FM layers was confirmed by X-ray photoelectron spectroscopic (XPS) measurements. The chemical states of Co, Ni, and Fe were detected by using a MgK$\rm\alpha$ (1253.6 eV) source and a hemispherical energy analyzer (pass energy of 10 eV with a resolution of ~0.03 eV). The XPS spectra were recorded following the procedure. Firstly, the chemical states of all the elements at the FM surface were examined after the natural oxidation. Then, the oxidized surface layer was removed by Ar ion milling in the XPS vacuum chamber, allowing us to observe the unoxidized FM bulk. As shown in Fig~\ref{fig2}(b), both the 2$p$$_{3/2}$ and 2$p$$_{1/2}$ peak positions of the naturally-oxidized surface Co shift to higher binding energy, i.e., 780.00 and 796.02 eV, respectively, compared with that of the bulk, 777.69 and 792.84 eV. Besides, the energy separation of the 2$p$ peaks are enhanced from $\Delta_{\rm Co}=15.20\ \rm eV $ to $\Delta_{{\rm CoO}_{x}}=16.02\ \rm eV $ with the appearance of the satellite peaks. These results provide the evidence of the oxidized Co surface~\cite{BIESINGER20112717,doi:10.1002/sia.740030412}. For Ni in the Py film, similar results were observed in the XPS spectra, shown in Fig~\ref{fig2}(c), indicating the surface oxidation of Ni. For Fe in the Py film, the shifts of 2$p$ peaks to high binding energy and enhancement of the energy separation of the two peaks in Fig~\ref{fig2}(d) confirm the formation of FeO$_{x}$~\cite{doi:10.1002/sia.740030412}, which is consistent with the scenario that Fe is preferentially oxidized in Py alloys~\cite{Nagai_1987}. These results show that the CoO$_{x}$ and PyO$_{x}$ layers were formed at the FM surface, while the bulk of the Co and Py remains unoxidized.

\begin{figure}[tb]
\includegraphics[scale=1.2]{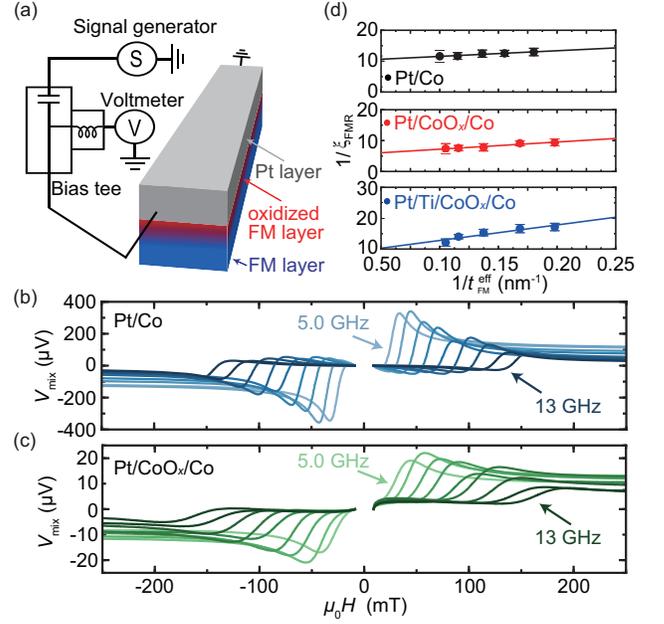}
\caption{(a) Schematic of the device structure and experimental set up for the ST-FMR measurement. The ST-FMR spectra $V_{\rm mix}$ measured at the frequency from $f= 5.0$ GHz to 13.0 GHz for the (b) Pt/Co and (c) Pt/CoO$_{x}$/Co films, respectively. The applied microwave power was 24.7 dBm. (d) The inverse of the ST-FMR spin-torque efficiency $1/\xi_{\rm FMR}$ as a function of the inverse of the effective Co layer thickness $1/t_{\rm FM}^{\rm eff}$ for the Pt/Co, Pt/CoO$_{x}$/Co, and Pt/Ti/CoO$_{x}$/Co films. The solid circles are the experimental data. The solid lines are the linear fitting results.}
\label{fig3} 
\end{figure}

To measure the SOTs using spin-torque ferromagnetic resonance (ST-FMR), the Pt/Co, Pt/CoO$_x$/Co, Pt/Py, and Pt/PyO$_x$/Py films were patterned into rectangular strips with the width of 10 $\mu$m and length of 100 $\mu$m using photolithography and Ar-ion etching. For the ST-FMR measurements, an in-plane external field $H$ was applied at an angle of $45^\circ$ from the longitudinal direction of the strips. A radio frequency (RF) current with a frequency of $f$ was also applied as shown in Fig~\ref{fig3}(a). The RF current in the device generates the DL- and FL-SOTs, as well as an Oersted field torque. The SOTs and Oersted field torque excite magnetic precession under the FMR condition: $(2\pi f/\gamma) = \sqrt{\mu_{0}H_\text{FMR}(\mu_{0}H_\text{FMR} + \mu_{0}M_\text{eff})}$, where $\gamma$ is the gyromagnetic ratio, $H_\text{FMR}$ is the FMR field, and $\mu_{0}M_\text{eff}$ is the effective demagnetization field~\cite{PhysRevB.92.064426}. The precessing magnetization generates a direct-current (DC) voltage $V_\text{mix}$ through the mixing of the applied RF current and oscillating resistance due to the anisotropic magnetoresistance (AMR) of the FM layer~\cite{liu2011spin,zhang2015role}:
\begin{eqnarray}
	\nonumber
	V_\text{mix}=V_\text{sym}\frac{W^2}{(\mu_0H-\mu_0H_\text{FMR})^2+W^2}\\
	+V_\text{anti}\frac{W(\mu_0H-\mu_0H_\text{FMR})}{(\mu_0H-\mu_0H_\text{FMR})^2+W^2}, \label{fitfunction}
\end{eqnarray}
where $W$ is the spectral width. $V_\text{sym}$ and $V_\text{anti}$ are the symmetric and antisymmetric components of $V_\text{mix}$. $V_\text{sym}$ is generated by the DL effective field $H_\text{DL}$ and $V_\text{anti}$ is generated by the FL effective field $H_\text{FL}$, as well as the Oersted field $H_\text{Oe}$~\cite{PhysRevB.92.064426}.

\section{III. results and discussion}

First, we study the effect of the interface oxidation on the SOTs in the Pt/Co bilayer. Figures~\ref{fig3}(b) and \ref{fig3}(c) show the DC voltage $V_\text{mix}$ measured for the Pt/Co and Pt/CoO$_{x}$/Co films at $f$ from 5 to 13 GHz. This result shows that both symmetric and antisymmetric voltage are generated in the Pt/Co and Pt/CoO$_{x}$/Co films. Here, by fitting the measured $V_\text{mix}$ spectra using Eq.~(\ref{fitfunction}), the FMR spin-torque generation efficiency, defined as 
\begin{equation}
\xi_\text{FMR}=\frac{{V_\text{sym}}}{{V_\text{anti}}}\frac{{e\mu_{0}M_\text{s}t^{\rm eff}_\text{FM} t_\text{Pt}}}{{\hbar}}\sqrt{1+\frac{{M_\text{eff}}}{H_\text{FMR}}},
\end{equation}
can be obtained, where $t_\text{Pt}$ is the thickness of Pt layer. $M_\text{s}$ is the saturation magnetization of the FM layer. 
The effective thicknesses of the FM layer $t^{\rm eff}_\text{FM}$ was determined from $t_{\rm FM}$ dependence of the saturation magnetization per unit area, measured using a vibrating sample magnetometer.  
The FMR $\xi_\text{FMR}$ spin-torque efficiency is related to the DL- and FL-SOT efficiencies, $
\xi_\text{DL(FL)} = (2e/\hbar )\mu_{0}M_\text{s}t^{\rm eff}_\text{FM}H_\text{DL(FL)}/j_{\rm Pt}$, where $j_{\rm Pt}$ represents the RF charge current density in the Pt layer, as~\cite{PhysRevB.92.064426}
\begin{equation}
\frac{1}{\xi_{\mathrm{FMR}}}=\frac{1}{\xi_{\mathrm{DL}}}\left(1+\frac{\hbar}{e} \frac{\xi_{\mathrm{FL}}}{4 \pi M_\text{s} t^{\rm eff}_{\mathrm{FM}} t_{\mathrm{Pt}}}\right).   \label{thicknesseq}
\end{equation}
Equation~(\ref{thicknesseq}) shows that the DL- and FL-SOT efficiencies, $\xi_\text{DL(FL)} $, can be determined by measuring $1/t^{\rm eff}$ dependence of $1/\xi_{\mathrm{FMR}}$. In Fig.~\ref{fig3}(d), we show the $1/t^{\rm eff}_\text{FM}$ dependence of $1/\xi_\text{FMR}$. From this result with Eq.~(\ref{thicknesseq}), we obtained the DL- and FL-SOT efficiencies for the Pt/Co film as $\xi_\text{DL}= 0.103$ and $\xi_\text{FL}=0.0303$. These values are consistent with $\xi_\text{DL(FL)}$ reported previously~\cite{PhysRevLett.116.126601}.

The SOT efficiencies in the Pt/Co bilayer are enhanced by the  oxidation of the interfacial Co layer. From the $1/t^{\rm eff}_\text{FM}$ dependence of $1/\xi_\text{FMR}$ for the Pt/CoO$_x$/Co film shown in Fig.~\ref{fig3}(d), we obtained the DL- and FL-SOT efficiencies as $\xi_\text{DL}= 0.204$ and $\xi_\text{FL}=0.0639$ for the Pt/CoO$_x$/Co film. The SOT efficiencies are summarized in Table I. This result shows that both DL and FL torque efficiencies are enhanced by the interfacial oxidation of the Pt/Co film, which is consistent with a previous report~\cite{PhysRevB.98.020405}. The previous study suggests that the enhancement of the SOTs is induced by the increase of the transmission efficiency of the spin current generated by the spin Hall effect in the Pt layer due to the antiferromagnetic nature of the CoO$_x$ layer. However, the origin of the enhancement, in particular the role of the interfacial spin-orbit coupling, remains unclear.

\begin{table}[] 
\caption{The DL-SOT efficiency $\xi_\text{DL}$ and the FL-SOT efficiency $\xi_\text{FL}$ for the Pt/Co, Pt/CoO$_x$/Co, Pt/Ti/CoO$_x$/Co, Pt/Py, and Pt/PyO$_x$/Py films. }
\begin{tabular}{lccccccc}
\hline \hline
 Layer structure &  $\xi_\text{DL}$  &  $\xi_\text{FL}$ \\
 \hline
Pt/Co & $0.103\pm0.018$ & $0.0303\pm0.017$  \\ 
Pt/CoO$_{x}$/Co & $0.204\pm0.022$ & $0.0639\pm0.013$ \\ 
Pt/Ti/CoO$_{x}$/Co & $0.116\pm0.015$ & $0.0478\pm0.014$  \\
Pt/Py & $0.0476\pm 0.0012$ & $0.00334\pm0.00104$ \\
Pt/PyO$_{x}$/Py & $0.00800\pm0.00068$ & $-0.0130\pm0.0024$ \\
\hline \hline
\end{tabular}
\end{table}

There are two possible sources responsible for the enhancement of the SOTs induced by the interfacial oxidation of the Pt/Co film: the bulk spin-orbit coupling and interfacial spin-orbit coupling. The DL efficiency due to the bulk spin-orbit coupling is expressed as $\xi_\mathrm{DL,bulk}=T_\mathrm{int}\sigma_\mathrm{SHE}$, where $T_\mathrm{int}(\leq 1)$ is the transparency of the spin current generated by the bulk spin Hall effect and $\sigma_\mathrm{SHE}$ is the spin Hall conductivity of the Pt layer. In the Pt/Co bilayer, the SOTs are dominated by this bulk mechanism and the SOTs arising from the interface are negligible~\cite{PhysRevLett.122.077201} . However,  $\xi_\mathrm{DL,bulk}$ can still be enhanced by the interfacial oxidation. The reason for this is that $T_\mathrm{int}$ can still be improved because at the non-oxidized Pt/Co interface, $T_\mathrm{int}$ is reduced by the spin memory loss, caused by the spin-orbit coupling. Thus, a possible origin of the enhancement of $T_\mathrm{int}$ that can be induced by the interfacial oxidation is the suppression of the interface spin memory loss, which can be induced when the interfacial oxidation suppresses the interfacial spin-orbit coupling. Another possible origin is the efficient spin transmission across the CoO$_x$ layer with the antiferromagnetic nature, produced by the interfacial oxidation.

To clarify the role of $T_\mathrm{int}$ in the enhancement of the SOT efficiencies induced by the interfacial oxidation, we measured the ST-FMR for the Pt/Ti/CoO$_x$/Co film with various $t^{\rm eff}_{\rm Co}$, where the 1-nm-thick Ti layer is inserted between the Pt and CoO$_x$ layers. Here, the thickness of the Ti layer is much smaller than the spin diffusion length of Ti, 13~nm~\cite{PhysRevB.90.140407}. From the $1/t^{\rm eff}_\text{FM}$ dependence of $1/\xi_\text{FMR}$ shown in Fig.~\ref{fig3}(d), we obtained the DL- and FL-SOT efficiencies $\xi_\text{DL}= 0.116$ and $\xi_\text{FL}=0.0478$ for the Pt/Ti/CoO$_x$/Co film; both DL and FL efficiencies in the Pt/CoO$_x$/Co film are clearly suppressed by the Ti insertion.

The suppression of $\xi_\mathrm{DL}$ induced by the Ti insertion indicates that the possible enhancement of $T_\mathrm{int}$ is not responsible for the enhancement of the SOTs induced by the interfacial oxidation. Here, $T_\mathrm{int}$ in the Pt/Ti/CoO$_x$/Co film is expected to be comparable or larger than that in the Pt/CoO$_x$/Co film because the possible efficient spin transport across the CoO$_x$ layer is present in both films, and the insertion of a thin layer with weak spin-orbit coupling, i.e., the Ti layer, reduces the interface spin memory loss, and therefore enhances $T_\mathrm{int}$~\cite{0034-4885-78-12-124501}. Thus, if $T_\mathrm{int}$ of the Pt/Co film is enhanced by the interfacial oxidation, and this enhancement is responsible for the enhancement of the SOT efficiencies, the same effect should be observed in the Pt/Ti/CoO$_x$/Co film, which is different from the experimental result. This shows that the possible enhancement of the SOT arising from the bulk spin Hall effect thorough the change of $T_\mathrm{int}$ is not the dominant mechanism of the enhancement of the SOT efficiencies. The minor role of the bulk spin Hall effect indicates that the spin-orbit coupling at the Pt/CoO$_x$ and/or CoO$_x$/Co interfaces is the source of the efficient SOTs in the Pt/CoO$_x$/Co film. The clear difference in $\xi_\mathrm{DL}$ between the Pt/CoO$_x$/Co and Pt/Ti/CoO$_x$/Co films shows that the Pt/CoO$_x$ interface is the main source of the efficient SOT generation, indicating a strong spin-orbit coupling at the Pt/CoO$_x$ interface.

\begin{figure}[tb]
\includegraphics[scale=1.2]{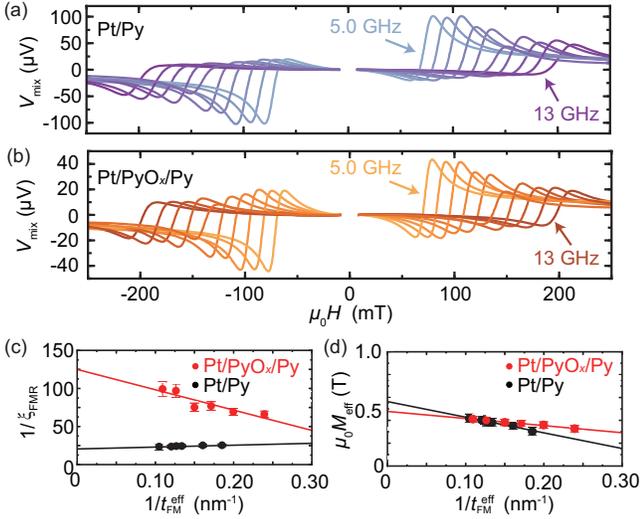}
\caption{The ST-FMR spectra $V_{\rm mix}$ measured at the frequencies from $f=5.0$ GHz to 13.0 GHz for the (a) Pt/Ni$_{81}$Fe$_{19}$ (Pt/Py) and (b) Pt/PyO$_{x}$/Py films, respectively. The applied microwave power was 24.7 dBm. (c) The inverse of the ST-FMR spin-torque efficiency $1/\xi_{\rm FMR}$ as a function of the inverse of the effective Py layer thickness $1/t_{\rm FM}^{\rm eff}$ for the Pt/Py and Pt/PyO$_{x}$/Py films. The solid circles are the experimental data. The solid lines are the linear fitting results. (d) The $1/t_{\rm FM}^{\rm eff}$ dependence of the effective demagnetization field $\mu_0 M_{\rm eff}$ for the Pt/Py and Pt/PyO$_{x}$/Py films. The solid circles are the experimental data. The solid lines are the linear fitting results.
}
\label{fig4} 
\end{figure}

The situation changes drastically by replacing the Co layer with Py. Figures~\ref{fig4}(a) and \ref{fig4}(b) show the ST-FMR for the Pt/Py and Pt/PyO$_{x}$/Py films. In Fig.~\ref{fig4}(c), we show the $1/t^{\rm eff}_{\rm FM}$ dependence of $1/\xi_\text{FMR}$. The $1/t^{\rm eff}_{\rm FM}$ dependence of $1/\xi_\text{FMR}$ shows that the intercept for the Pt/PyO$_x$/Py film is clearly larger than that for the Pt/Py film, indicating smaller $\xi_\mathrm{DL} $ in the Pt/PyO$_x$/Py film. This result further shows that the sign of the slope of the $1/t^{\rm eff}_{\rm FM}$ dependence of $1/\xi_\text{FMR}$ is reversed by the interfacial oxidation, showing that the sign of $\xi_\mathrm{FL}$ is opposite between the Pt/Py and Pt/PyO$_x$/Py films. By fitting the result shown in Fig.~\ref{fig4}(c), we obtained the SOT efficiencies $\xi_\text{DL}$ and $\xi_\text{FL}$ for the Pt/Py and Pt/PyO$_{x}$/Py films, as shown in Table~I.

In the Pt/Py film, the interfacial oxidation suppresses $\xi_\text{DL}$ from 0.0476 to 0.00800, which is in contrast to the result for the Pt/Co film, where the interfacial oxidation enhances $\xi_\text{DL}$. The DL-SOT in the Pt/Py bilayer is dominated by the bulk spin Hall effect of the Pt layer~\cite{PhysRevB.91.214416}, indicating that the suppression of $\xi_\text{DL}$ arises from the suppression of $T_\mathrm{int}$, rather than the suppression of the DL-SOT originating from the interface. There are two possibilities that suppress $T_\mathrm{int}$. A possibility is the enhancement of the interface spin memory loss due to the interfacial oxidation. Another possibility is that the PyO$_x$ layer suppresses the real part of the bare spin mixing conductance $\text{Re}[G^{\uparrow\downarrow}]$ due to the insulating nature~\cite{doi:10.1063/1.3280378}.

The suppression of $\xi_\text{DL}$ in the Pt/Py film cannot be attributed to the change of the interface spin memory loss. The evidence for this was obtained from the magnetic damping affected by the spin pumping. In the Pt/Py bilayer, the spin pumping driven by FMR emits a spin current from the Py layer into the Pt layer, which gives rise to additional magnetic damping. The effective magnetic damping $\alpha_{\rm{eff}}$ in the presence of the spin pumping is expressed as~\cite{PhysRevLett.123.057203}
\begin{align}
\alpha_{\rm{eff}}=\alpha_{\text{int}}+\text{Re}[G_{\rm{eff,tot}}^{\uparrow\downarrow}]\frac{g\mu_{\text{B}} h}{4\pi e^2 M_{\text{s}}}\frac{1}{d_{\text{FM}}},
\label{G_eff-thick}
\end{align} 
where $\alpha_{\rm{int}}$ is the intrinsic magnetic damping of the Py layer, $g$ is the $g$-factor, $\mu_{\rm{B}}$ is the Bohr magnetron, and $h$ is the Planck constant. The real part of a total effective spin mixing conductance $\text{Re}[G_{\text{eff,tot}}^{\uparrow\downarrow}]$ consists of two components: $\text{Re}[G_{\text{eff,tot}}^{\uparrow\downarrow}]=\text{Re}[G_{\text{eff}}^{\uparrow\downarrow}]+G_{\text{SML}}$, where $G_{\text{eff}}^{\uparrow\downarrow}$ is the effective spin mixing conductance and $G_{\rm{SML}}$ is the additional component due to the spin memory loss at the Pt/FM interface. $\text{Re}[G_{\rm{eff,tot}}^{\uparrow\downarrow}]$ for the Pt/Py and Pt/PyO$_x$/Py films can be extracted by fitting $1/d_{\rm{FM}}$ dependence of $\alpha_{\rm{eff}}$ using Eq.~(\ref{G_eff-thick}). The result is $\text{Re}[G_{\rm{eff,tot}}^{\uparrow\downarrow}]=1.14 \times 10^{19} \rm m^{-2}$ for the Pt/Py film and $\text{Re}[G_{\rm{eff,tot}}^{\uparrow\downarrow}]=0.567 \times 10^{19} \rm m^{-2}$ for the Pt/PyO$_x$/Py film, showing that $\text{Re}[G_{\rm{eff,tot}}^{\uparrow\downarrow}]$ is suppressed by the interfacial oxidation. This indicates that $G_{\rm{SML}}$ at the Pt/PyO$_x$ and PyO$_x$/Py interfaces is not significant, showing that the possible enhancement of the spin memory loss is not the dominant source of the suppression of $\xi_\text{DL}$ in the Pt/PyO$_x$/Py film. The minor role of the change of the interface spin memory loss is supported by the interface perpendicular magnetic anisotropy (PMA) energy density $K_{\rm s}$, which originates from the interfacial spin-orbit coupling~\cite{PhysRevLett.81.5229,PhysRevB.72.054430}. For the Pt/Py and Pt/PyO$_x$/Py films, $K_{\rm s}$ was determined from $1/t_{\rm FM}$ dependence of the effective demagnetization field $\mu_{0} M_{\rm eff}$, shown in Fig.~\ref{fig4}(d), using
\begin{equation}
\mu_0 M_{\rm eff} = \mu_0 M_{\rm s} -\frac{2K_{\rm s}}{M_{\rm s}t^{\rm eff}_{\rm FM}},  \label{Meff}
\end{equation}
where $\mu_{0} M_{\rm eff}$ was obtained by fitting $f$ dependence of $H_\mathrm{FMR}$ using Kittel formula. The obtained interface PMA energy density is $K_{\rm s}=0.308$ mJ/m$^2$ for the Pt/Py film and $K_{\rm s}=0.118$ mJ/m$^2$ for the Pt/PyO$_x$/Py film; $K_{\rm s}$ is decreased by the interfacial oxidation. This result suggest that the interfacial spin-orbit coupling and interface spin memory loss are suppressed by the interfacial oxidation, since a previous study has shown that the increase of the interface spin memory loss due to the spin-orbit coupling at HM/FM interfaces is associated with the increase of $K_{\rm s}$~\cite{PhysRevLett.122.077201}. 
These results indicate that the suppression of $\xi_\mathrm{DL}$ is originates from the suppression of the real part of the bare spin mixing conductance $\text{Re}[G^{\uparrow\downarrow}]$, which suppresses the interface spin transparency $T_\mathrm{int}$.

The interfacial oxidation of the Pt/Py bilayer changes the sign and enhances the magnitude of $\xi_\mathrm{FL}$, as shown in Table I, which can be attributed to the change of the imaginary part of the spin mixing conductance $\mathrm{Im}[G^{\uparrow \downarrow}]$. The minor role of the spin memory loss in the Pt/PtO$_x$/Py film shows that the interfacial spin-orbit coupling is not significant, suggesting that the FL-SOT in this film mainly originates from the bulk spin Hall effect in the Pt layer. 
In the scenario of the bulk spin Hall effect, the DL-SOT is generated by the injection of the spin current into the ferromagnetic layer, while the FL-SOT arises from the reflection of the spin current at the interface. The DL-SOT and FL-SOT in this model are approximately proportional to the real and imaginary parts of the spin mixing conductance, $\mathrm{Re}[G^{\uparrow \downarrow}]$ and $\mathrm{Im}[G^{\uparrow \downarrow}]$, respectively~\cite{PhysRevB.94.104420}. The insulating PyO$_x$ layer suppresses the real part of the spin mixing conductance $\mathrm{Re}[G^{\uparrow \downarrow}]$~\cite{doi:10.1063/1.3280378}, resulting in the suppression of $\xi_\mathrm{DL}$ by the interfacial oxidation. In contrast, $\mathrm{Im}[G^{\uparrow \downarrow}]$ can be enhanced due to the barrier because a spin current reflected at the interface experiences a large angle rotation of its spin direction, which corresponds to a larger value of $\mathrm{Im}[G^{\uparrow \downarrow}]$. This is consistent with the enhancement of $\xi_\mathrm{FL}$. We also note that the sign of $\mathrm{Im}[G^{\uparrow \downarrow}]$ is sensitive to the electronic structure of the interface; the sign of $\xi_\mathrm{FL}$ is opposite due to the opposite sign of $\mathrm{Im}[G^{\uparrow \downarrow}]$ in Pt/Ni and Pt/Fe bilayers~\cite{hayashi2020spinorbit}. This suggests that the interfacial oxidation of the Pt/Py bilayer changes the magnitude and sign of the imaginary part of the spin mixing conductance, resulting in the change of the FL-SOT originating from the bulk of the Pt layer.

\section{IV. conclusion}
In summary, we investigated the DL- and FL-SOT efficiencies in the HM/FM structures in the presence of the interfacial oxidation. In the Pt/Co bilayer, the interfacial oxidation enhances both DL- and FL-SOT efficiencies. The dominant source of the SOTs in the Pt/CoO$_x$/Co film is the spin-orbit coupling at the Pt/CoO$_x$ interface, showing that the interfacial oxidation enhances the interfacial spin-orbit coupling. In contrast, in the Pt/Py film, the interfacial oxidation suppresses the DL-SOT efficiency, while it changes the sign of the FL-SOT. In the Pt/Py and Pt/PyO$_x$/Py films, the interfacial spin-orbit coupling plays a minor role, as evidenced by the effective spin mixing conductance and interface magnetic anisotropy. In the Py-based films, the bulk spin-orbit coupling is responsible for the SOTs. In this system, the change of the SOTs can be attributed to the change of the real $\mathrm{Re}[G^{\uparrow \downarrow}]$ and imaginary $\mathrm{Im}[G^{\uparrow \downarrow}]$ parts of the spin mixing conductance. The interfacial oxidation suppresses $\mathrm{Re}[G^{\uparrow \downarrow}]$, which is associated with the enhancement of $\mathrm{Im}[G^{\uparrow \downarrow}]$, due to the insulating nature of the PyO$_x$ layer. These results demonstrate that the interfacial oxidation provides an effective way to manipulate the strength and sign of the SOTs.

\begin{acknowledgments}
This work was supported by JSPS KAKENHI Grant Numbers 19H00864, 19K22131, Canon Foundation, Asahi Glass Foundation, Kao Foundation for Arts and Sciences, JGC-S Scholarship Foundation, and Spintronics Research Network of Japan (Spin-RNJ). 
\end{acknowledgments}



%

\end{document}